# *Multimodal Medical Image registration using Discrete Wavelet Transform and Gaussian Pyramids*

Engr. Hina Shakir
Electrical Engineering Department
Bahria University
Karachi, Pakistan
hina.shakir@bimcs.edu.pk

Dr. S. Talha Ahsan
Electrical Engineering Department
Usman Institute of Technology
Karachi, Pakistan
stahsan@uit.edu

Engr. Nabiha Faisal
Computer & Software Engineering Department
Bahria University
Karachi, Pakistan
nabiha_faisal@bimcs.edu.pk

***Abstract*—In this research paper, authors propose multimodal brain image registration using discrete wavelet transform(DWT) followed by Gaussian pyramids. The reference and target images are decomposed into their LL, LH, HL and LL DWT coefficients and then are processed for image registration using Gaussian pyramids. The image registration is also done using Gaussian pyramids only and wavelets transforms only for comparison. The quality of registration is measured by comparing the maximum MI values used by the three methods and also by comparing their correlation coefficients. Our proposed technique proves to show better results when compared with the other two methods.**

***Keywords—multimodal; discrete wavelet transform; gaussian pyramids; image registration; medical images***

## I. INTRODUCTION

Multimodal medical image registration establishes an association between two medical images of the anatomically identical structures captured from different devices at different time frames from different perspectives. It aligns two images known as reference and target images with the help of some transformation and has proven to be very beneficial in medical image analysis processes like tumor growth, localizing lesions or deformation of an organ or tissue.

Rigid body image registration requires an iterative process which begins from initial estimate position to reach an optimal solution. The optimization can stop at false points due to misguidance from local minima points while measuring intensity similarity of the registering images. Another important factor influencing process of registration is the capture range which is defined as a range of pixels data of the image within which the registration algorithm starts and converges to the correct values [1]. False registration can also take place if the algorithm starts at a point which does not fall in the capture range resulting in failure of the convergence of algorithm. Therefore larger size of capture range promises higher success rate of a registration.

The traditional rigid body image registration techniques use Gaussian pyramids more commonly known as Gaussian filters to register the images in spatial domain and employ mutual information as a similarity metric for optimization of the registration algorithm. However, the problem with Gaussian pyramids is short range of information captured in the image especially at low frequency resolutions during iteration process performed to achieve optimum results [2]. The size of capture range depends upon the intensity, features and similarity of the two images used. Discrete wavelet transform (DWT) is an efficient and powerful technique that has been used to obtain important geometric characteristics for registration [3-8]. DWT decomposes the image into its details and approximation coefficients which enables refined processing of the image based on the multi resolution data made available. In fact Image registration performed on wavelet coefficients of brain images is found to produce better results due to larger capture ranges as compared to the registration performed on Gaussian pyramids [8].

In this research paper, authors propose to increase the capture range by using discrete wavelets transform (DWT) along with Gaussian pyramids for better multimodal image registrations. The tested wavelet is Haar wavelet and the similarity metric used for registration is Matte’s mutual information. Mutual information (MI) is a widely used matching criterion for multimodal image registration [9-10] and the efficiency of MI can be increased by using multi resolution scheme.

The reference and target images are decomposed in their respective LL, HL, LH and HH frequency sub-bands using 2D DWT. The four frequency sub-bands are registered using

Gaussian pyramids of level 3. Finally, the registered data available in wavelet domain is inverse transformed using Inverse DWT into registered image.

For comparison, the images are also registered using Gaussian pyramid and wavelets domain separately. The performance of the proposed technique is measured by comparing the maximum value of mutual information used by all three registration processes and also by comparing the correlation coefficients of reference and registered images.

The proposed framework of image registration is outlined in Fig. 1.

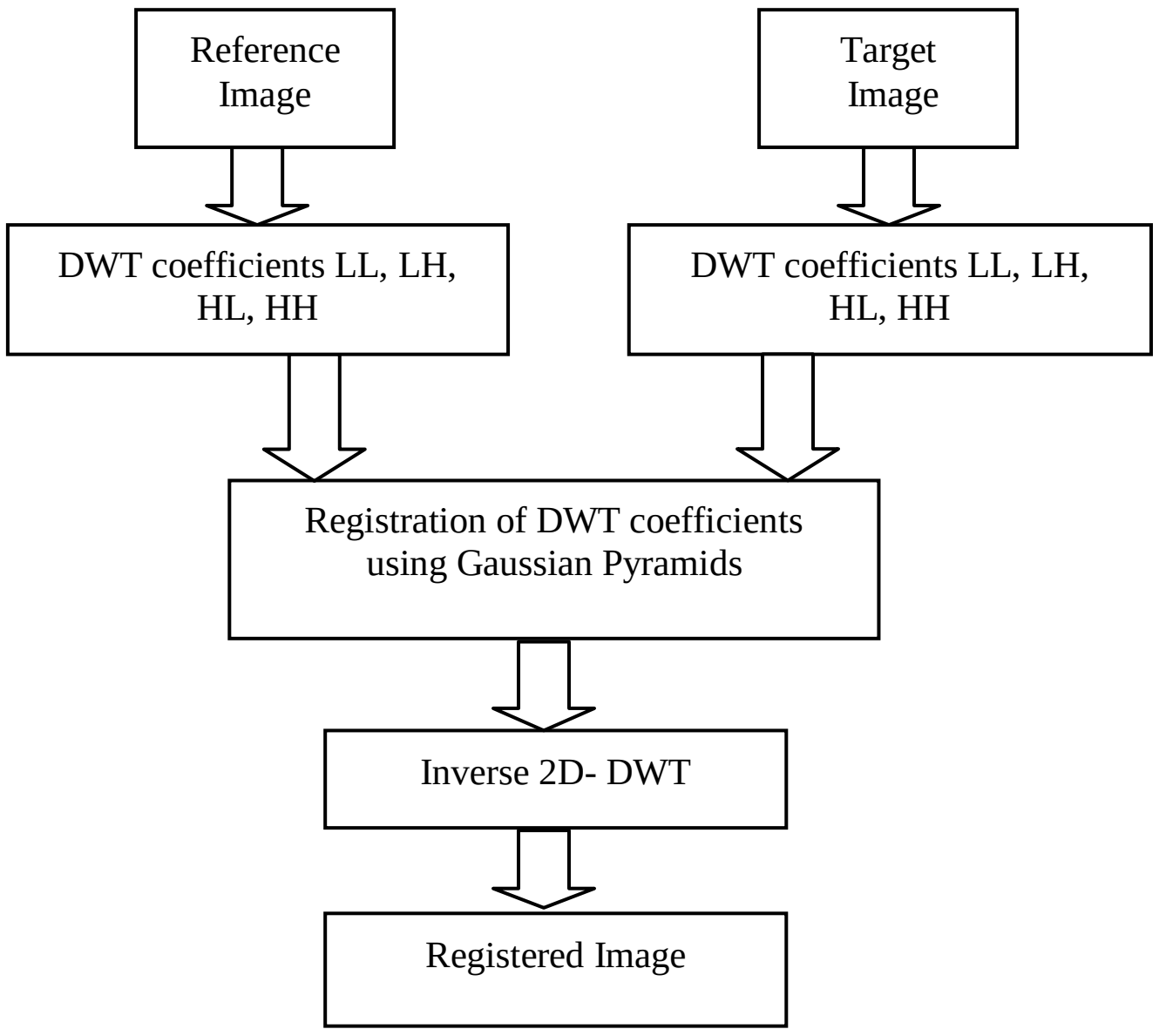

Fig. 1. The proposed model of image registration

## II. Background

This section briefly discusses wavelet transform, Gaussian pyramids and image registration process employed for our research work.

### *A. Wavelet Transfom*

The reference and the moving images are decomposed into their respective Haar wavelet coefficients using DWT. DWT scales time and frequency components of the image which helps in multi-level, multi-resolution analysis of the images.

Wavelets produce different results in both time and frequency domain for different ranges of frequencies. For low frequencies, the results are better in frequency plane where as for higher frequencies the performance of wavelets is superior in spatial domain. Wavelet transforms provide good image registration due to their multi-resolution nature.

### *B. Gaussian Pyramids*

Pyramids of hierarchical imaging are used to effectively perform the optimization. The proposed methodology uses Gaussian pyramids of 3 levels for registration of wavelet coefficients of reference and the moving image.

Gaussian pyramids make a pyramid of image information by capturing the data at different levels of the image and getting to the lower resolutions of the image as well to find mutual dependence of the data available for the two data sets. The Gaussian pyramid creates low-pass filtered (i.e., down-sampled) images with reduced density for each level of pyramid from the preceding level, where the base level of the pyramid is the original image itself [11]. The Gaussian pyramid can be mathematically defined as follows:

$G_0$ (x, y) = I(x, y), for level L = 0 where I(x, y) is the original image

$$G_l(x,y)=\sum_{m=-2}^{2}\sum_{n=-2}^{2} w(m,n)G_{l-1}(2x+m, 2y+n) \qquad (1)$$

Where w(m, n) is a weighting function (identical at all levels) termed as the generating kernel which adheres to the following properties: separable, symmetric and each node at level n contributes the same total weight to nodes at level L+1.

Fig. 2 shows level 3 Gaussian pyramids used by for optimization.

## III. Image Registration

The necessary aspects which have to be been defined for the image registration are transformation model, a similarity metric, and an optimization method [12].

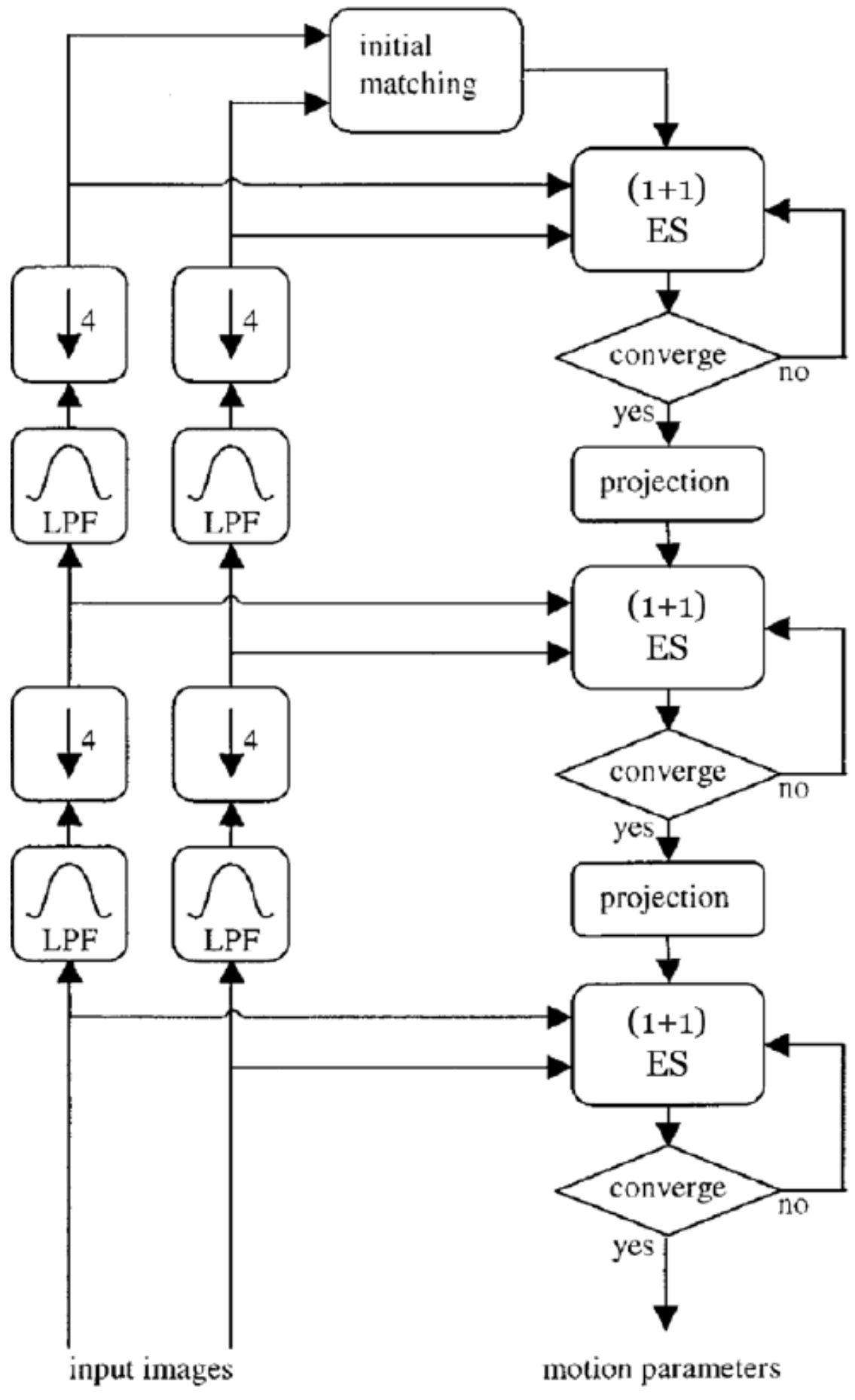

Fig. 2. Optimization algorithm for image registration

### A. *Affine Transformation Model*

For our rigid body image registration, affine transformation which is a combination of translation, rotation and scaling is used. Our research work employs MATLAB's affine transformation function that uses the same transformation method but with respect to the center pixel and with different variables and definitions for geometric parameters.

Let two registered data sets be **Z(u) and** $\mathbf{N^r(u)}$ sampled on $\mathbf{u} \in \mathbf{R}^2$ . **Z** is the fixed image defined on image coordinates **u** and **N** is the moving image defined on image coordinates **v.** $\mathbf{N^r}$ is the registered moving image. **Z (u)** and **N (v)** are observed. The relationship between $\mathbf{N^r}$ **(u)** and **N (v)** is given in (2) as:

$$\mathbf{N^r(u) = N(G(v)) \Leftrightarrow N(v) = N^r(G^{-1}(u))} \qquad (2)$$

Where **G** is the affine transformation
**G** is the product of 4 geometric transformations, translation **(**in x & y), Rotation (along x), Scaling (in x & y) and skew. Relationship between the transformations is described in (3).

$$B = \{t_x, t_y\, \theta, s_x, s_y, k\} \qquad (3)$$

Transformation parameters are:

$$g = [g_1, g_2, g_3, g_4, g_5, g_6]^T$$

$$u = \mathbf{G}.y \qquad (4)$$

$$G_B = \underbrace{\begin{bmatrix} 1 & 0 & t_x \\ 0 & 1 & t_y \\ 0 & 0 & 1 \end{bmatrix}}_{\textit{Translation}} \underbrace{\begin{bmatrix} \theta_c & -\theta_s & 0 \\ \theta_s & \theta_c & 0 \\ 0 & 0 & 1 \end{bmatrix}}_{\textit{rotation}} \underbrace{\begin{bmatrix} 1 & k & 0 \\ 0 & 1 & 0 \\ 0 & 0 & 1 \end{bmatrix}}_{\textit{skew}} \underbrace{\begin{bmatrix} s_x & 0 & 0 \\ 0 & s_y & 0 \\ 0 & 0 & 1 \end{bmatrix}}_{\textit{scaling}}$$

$$\begin{bmatrix} s_x\theta_c & s_y(k\theta_c - \theta_s) & t_x \\ s_x\theta_s & s_y(k\theta_s + \theta_c) & t_y \\ 0 & 0 & 1 \end{bmatrix} = \begin{bmatrix} g_1 & g_2 & g_3 \\ g_4 & g_5 & g_6 \\ 0 & 0 & 1 \end{bmatrix}$$

Where,
$\mathbf{t_x}$ **:** positive value shifts image to the left
$\mathbf{t_y}$ **:** positive value shifts image up
**θ :** rotation angle, measured counterclockwise from x-axis ($\theta_c = \cos\theta$ & $\theta_s = \sin\theta$ )
**k :** shear factor along the x-axis = tan(skew angle, measured from y-axis)
$\mathbf{s_x}$ **&** $\mathbf{s_y}$ **:** change of scale in x direction & y direction respectively

If the centre pixel is taken as $\mathbf{(x_i, y_i)}$

$$u = G.\, v_{(xi,yi)} = \begin{bmatrix} g_1 & g_2 \\ g_4 & g_5 \end{bmatrix} \begin{bmatrix} v_x - x_t \\ v_y - y_t \end{bmatrix} + \begin{bmatrix} g_3 + x_t \\ g_6 + y_t \end{bmatrix}$$

$$G_\beta = \begin{bmatrix} g_1 v_x + g_2 v_y + g_3 - g_1 x_t - g_2 y_t + x_t \\ g_4 v_x + g_5 v_y + g_6 - g_4 x_t - g_5 y_t + y_t \end{bmatrix}$$

Now,

$$u = \begin{bmatrix} h_1 & h_4 & 0 \\ h_2 & h_5 & 0 \\ h_3 & h_6 & 1 \end{bmatrix}^T v_{(1,1)} = \begin{bmatrix} g_1 & g_2 & g_3 - g_1 x_t - g_2 y_t + x_t \\ g_4 & g_5 & g_6 - g_4 x_t - g_5 y_t + y_t \\ 0 & 0 & 1 \end{bmatrix} \begin{bmatrix} v_x \\ v_y \\ 1 \end{bmatrix}$$

$$H = \begin{bmatrix} g_1 & g_4 & 0 \\ g_2 & g_5 & 0 \\ g_3 - g_1 x_t - g_2 y_t + x_t & g_6 - g_4 x_t - g_5 y_t + y_t & 1 \end{bmatrix}$$

$$H = \begin{bmatrix} h_1 & h_4 & 0 \\ h_2 & h_5 & 0 \\ h_3 & h_6 & 1 \end{bmatrix} =$$

$$\begin{bmatrix} s_x\theta_c & s_x\theta_s & 0 \\ s_y(k\theta_c - \theta_s) & s_y(k\theta_s + \theta_c) & 0 \\ t_x + x_t + s_y y_t(\theta_s - k\theta_c) - s_x\theta_c x_t & t_y + y_t - s_y y_t(k\theta_s + \theta_c) - s_x\theta_s x_t & 1 \end{bmatrix}$$

(5)

## B. Mutual Information

Mutual information(MI) is a metric which finds the mutual dependence of the two images with each other. Image registration maximizes the MI when the two images are in alignment with each other because of the information they share about each other. The MI equation used in our research is given in (6).

$$I(Z(u).N(G.v)) = \sum_{L \in L_N} \sum_{K \in L_Z} p_{ZN}(l,k;\ g)\ \log_2\left[\frac{p_{ZN}(l,k;\ g)}{p_Z(k).p_N(l;\ g)}\right] \quad (6)$$

Where $L_Z$ and $L_N$ are the intensities of the fixed and moving images, and $p_{ZN}$ , $p_Z$, and $p_N$ are the joint, fixed marginal, and moving marginal probability distributions, respectively.

A good optimization technique should work in synchronization of the similarity metric and should align the images to the optimum for registration having a given parameter set and transformation.

## C. Experiment Setup

To test our proposed methodoloy, we have accessed CT, PET, MR-T2 brain images from the Retrospective Image Registration Evaluation (RIRE) Project [13]. The chosen dataset is a set of brain images from five patients. The registration takes place between PET and MR-T2 and, CT and MR-T2 images for the available datasets of the patients.

The system configuration used for the experiment is 4 GB RAM with Itanium 5 processor. The proposed image registration technique is implemented using MATLAB 2014. The variables of optimizer and metric values employed in the MATLAB code are depicted in Table I with the following sets of values:

| Variable | Data type | Fields | Value |
|---|---|---|---|
| **Optimizer** | OnePlusOne Evolutionary | GrowthFactor | 1.01 |
| | | Epsilon | 1.5*e* – 6 |
| | | InitialRadius | 0.001 |
| | | MaximumIterations | 500 |
| **Metric** | MattesMutual Information | NumberOf SpatialSamples | 500 |
| | | NumberOf HistogramBins | 50 |
| | | UseAllPixels | 1 |

TABLE I. PARAMETERS USED BY MATLAB CODE FOR REGISTRATION

## D. Performance of proposed technique

The performance of proposed registration is evaluated using maximum MI and the correlation coefficient (CC) [14] obtained during each registration process. These two parameters are also calculated for wavelets and Gaussian pyramids separately.

Correlation $r_i$ between the reference and the registered image is measured using equation (7) [15]:

$$r_i = \frac{\sum_i (xi - xm)(yi - ym)}{\sqrt{\sum_i (xi - xm)^2}\sqrt{\sum_i (yi - ym)2}} \quad (7)$$

Where $x_i$ is the intensity of the $i^{th}$ pixel in registered image and $y_i$ is the intensity of the $i^{th}$ pixel in the fixed image. $x_m$ is the mean intensity of the registered image and $y_m$ is the mean intensity of the fixed image.

# IV. RESULTS

MR-T2 to PET and CT to MR-T2 image registration is performed for five patients and the performance parameters are recorded in Fig. 3 and Fig. 4 for spatial domain registration, wavelet based registration and wavelet followed by Gaussian pyramids based registration. MR-T2 and CT images during registration process of the proposed method are shown in Fig. 5. Common area between the reference and the registered image is shown in Fig. 5(d) using grey color where as Fuchsia color denotes difference between the two images.

The statistical performance parameters recorded in Fig. 3 and Fig. 4 reveal that our proposed technique outperforms spatial domain image registration using Gaussian pyramids and using wavelets both, showing the highest value of MI obtained for all the patients thus better registration. The correlation coefficient values of the input image and the registered image applying the three techniques for five patients are listed in Fig. 6 and Fig. 7. The correlation coefficient of our registration technique is 1 decimal place higher than the other two techniques.

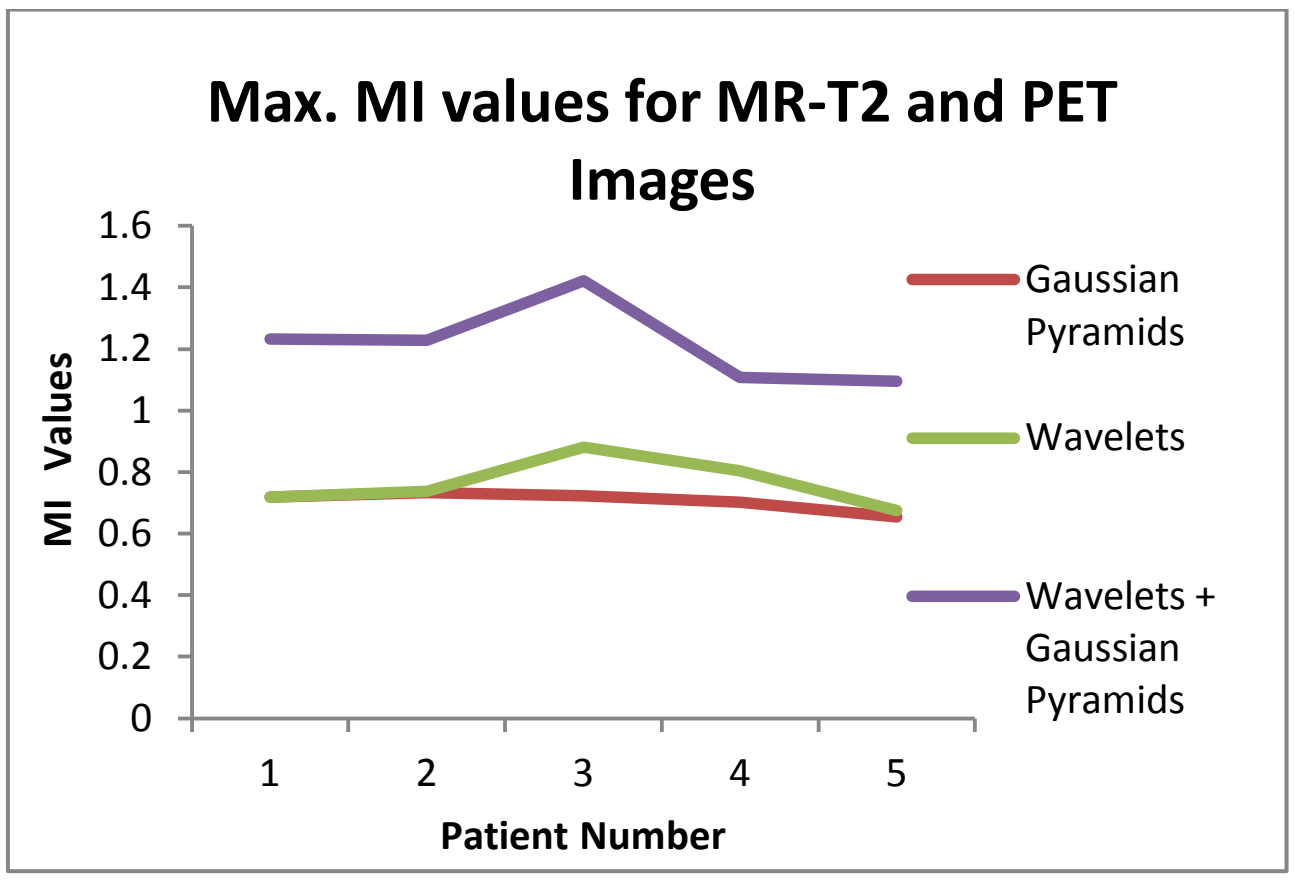

Fig. 3. Maximum values of MI obtained for the three methods on a set of five patients between MR-T2 and PET images

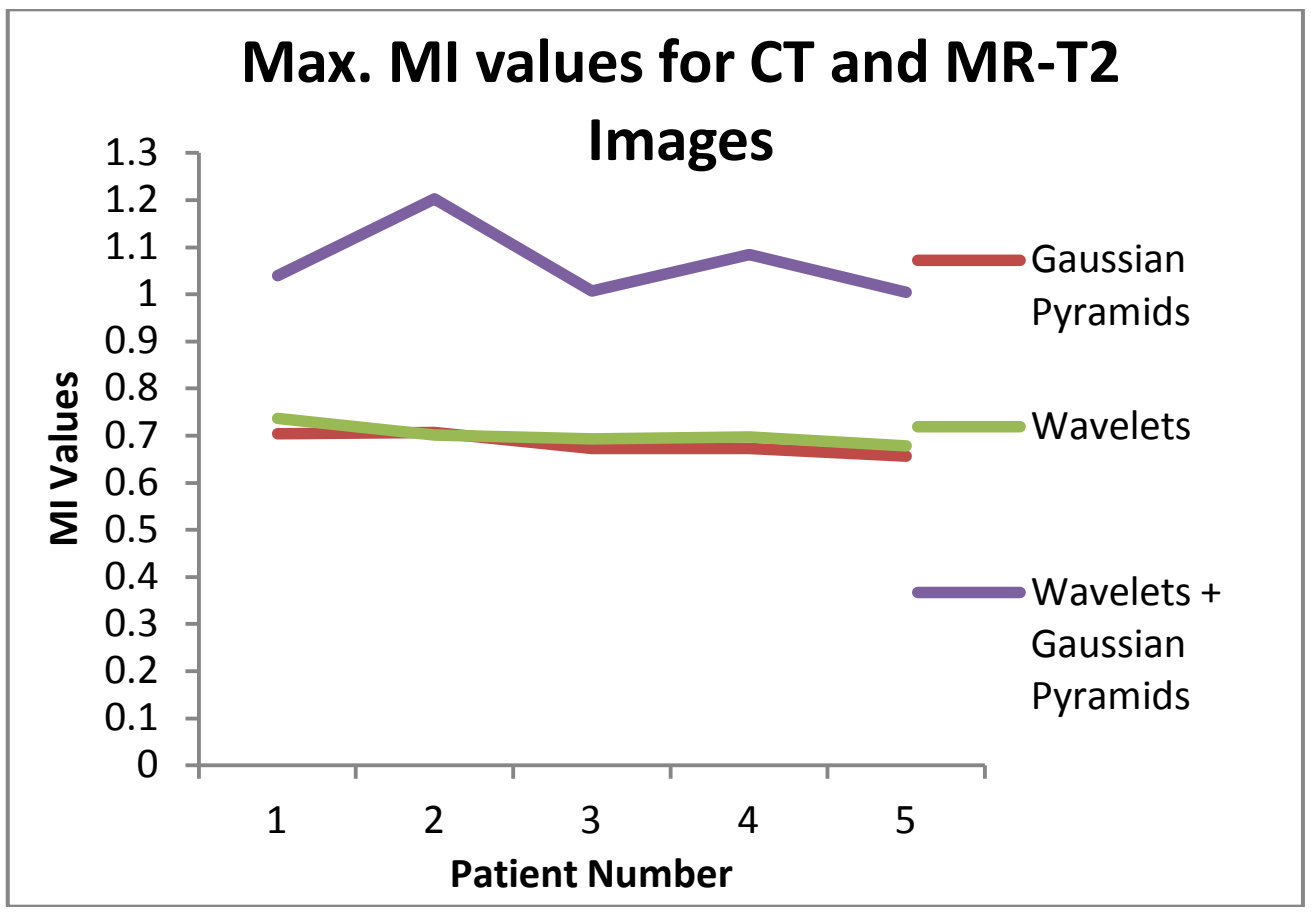

Fig. 4. Maximum of MI for the three methods on a set of five patients between CT and MR-T2 images

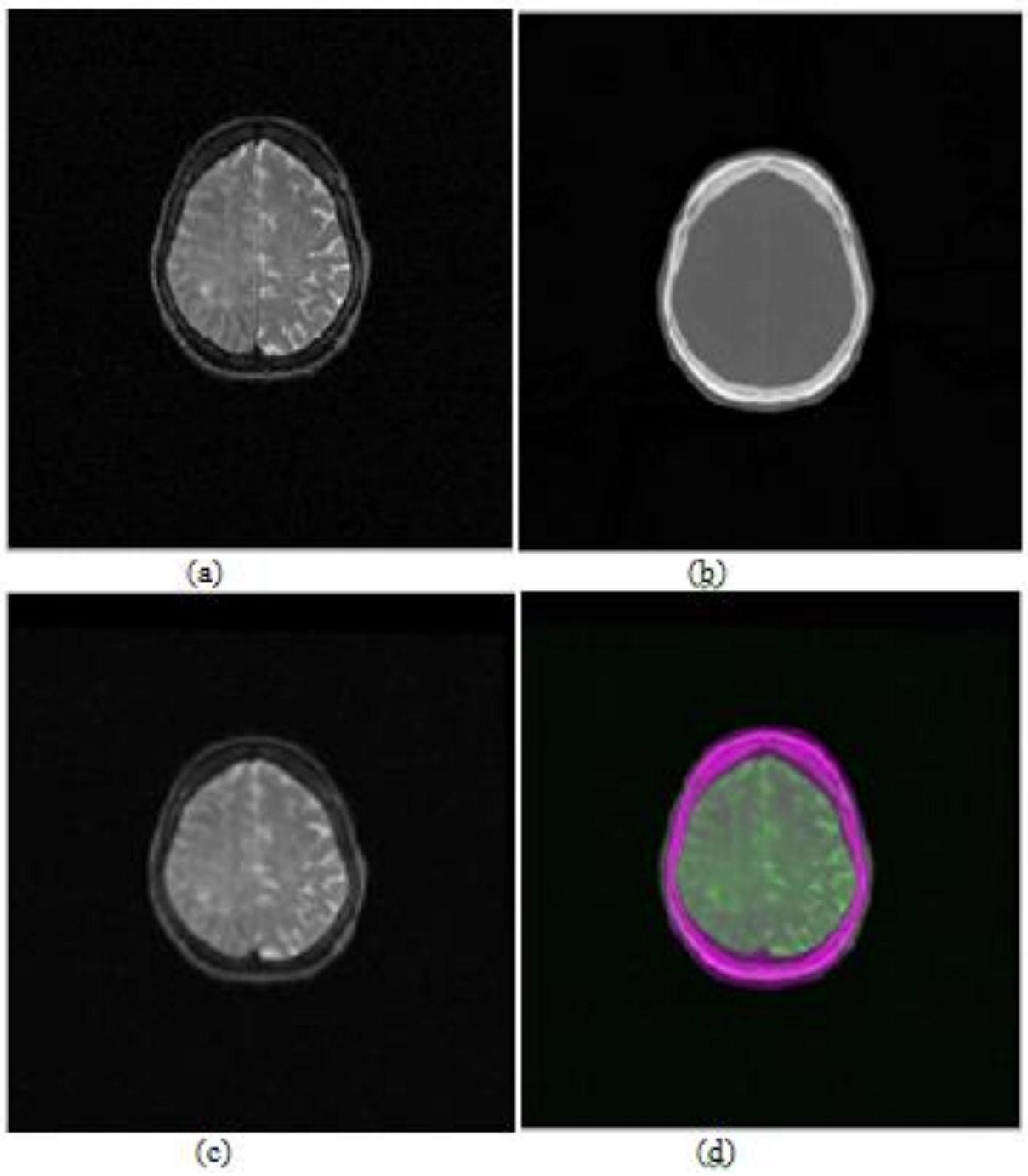

Fig. 5. Axial view of brain images of patient 01
(a) MR-T2 (reference) (b) CT image (target) (c) Registered image (d)Difference of registered and reference image

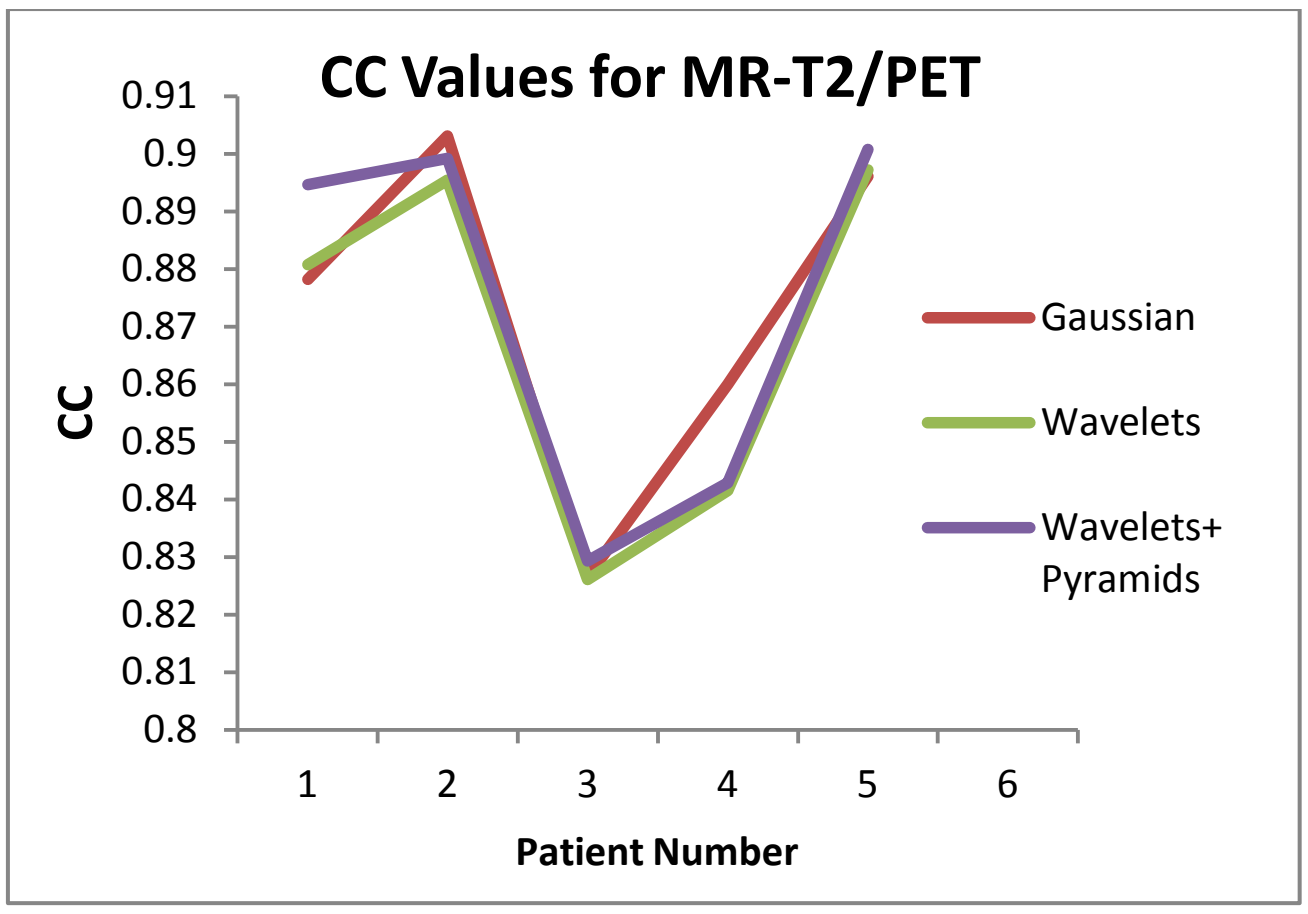

Fig. 6. Correlation Coefficients Between MR-T2 vs PET Samples of Patients for the three techniques

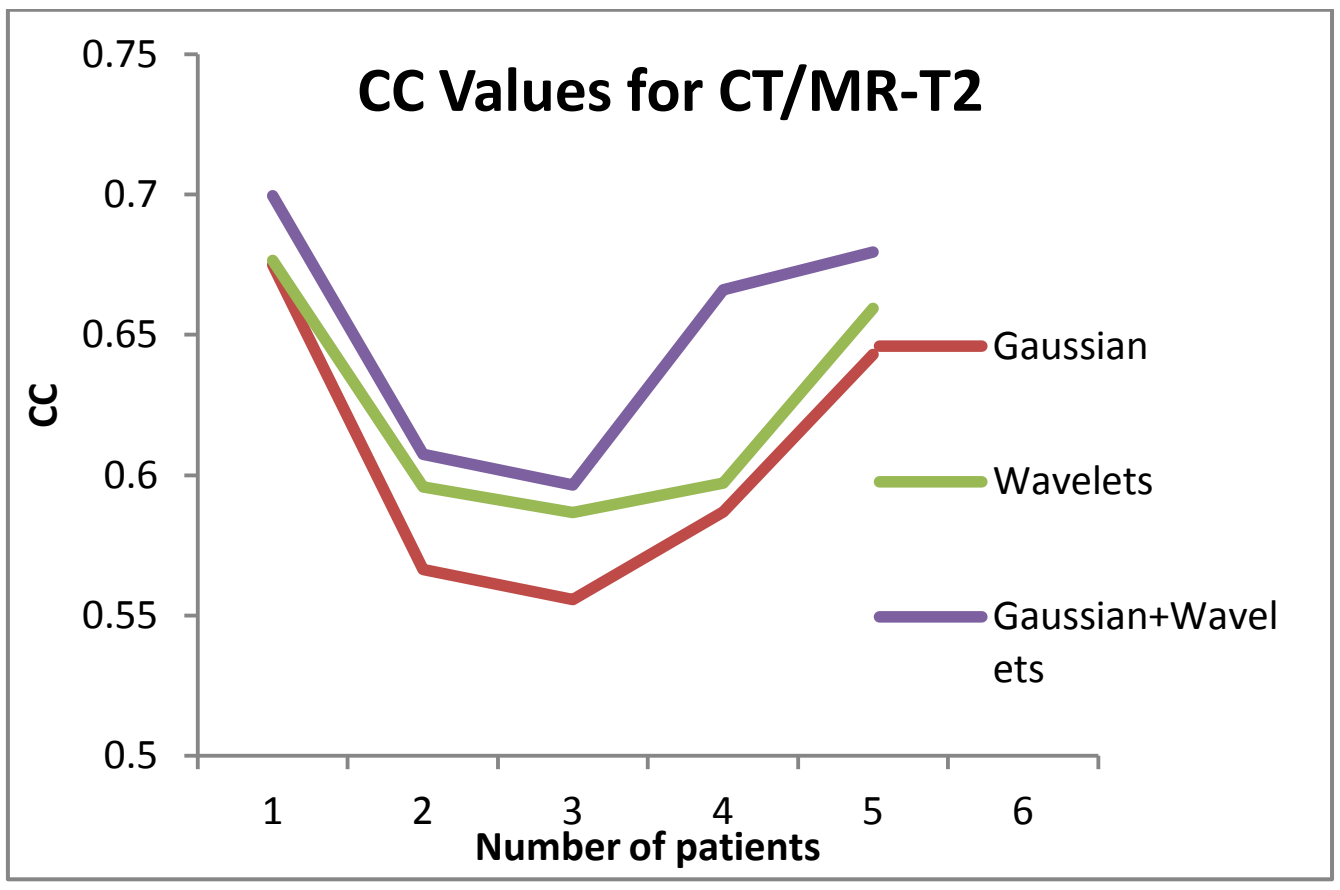

Fig. 7. Correlation Coefficients between CT vs MR-T2 Samples of Patients for the three methods

## V. CONCLUSION

This research paper proposes a novel technique to improve the quality of multimodal image registration algorithm. The wavelet coefficients of brain images have been registered using Gaussian pyramid and inverse transformed to obtain the registered image. The performance of the registration has been assessed using Mutual information and correlation coefficients. The proposed technique is compared with the two prevailing techniques of image registration and has proven to perform better in terms of MI and CC. Future research work can include testing of the proposed method on large database of patients and employing advance techniques to measure the performance of registration.